\documentclass[aps,prb,showpacs,reprint,superscriptaddress]{revtex4-1}

\usepackage{graphicx}
\usepackage{color}
\usepackage{amsmath}
\usepackage{bbm}
\usepackage{amssymb}

\usepackage{dsfont}

\usepackage[utf8]{inputenc}	
\usepackage[T1]{fontenc}

\input epsf

\begin{document}

\title{Superconducting dome with $extended$ $s$-$wave$ pairing symmetry in the heavily hole-overdoped copper-oxide planes}
\author{M. Zegrodnik}
\email{michal.zegrodnik@agh.edu.pl}
\affiliation{Academic Centre for Materials and Nanotechnology, AGH University of Science and Technology, Al. Mickiewicza 30, 30-059 Krakow,
Poland}
\author{P. W\'ojcik}
\affiliation{AGH University of Science and Technology, Faculty of Physics and Applied Computer cience, Al. Mickiewicza 30, 30-059 Krakow, Poland}
\author{J. Spa\l ek}
\affiliation{Institute of Theoretical Physics, Jagiellonian University, ul. \L ojasiewicza 11, 30-348 Krakow, Poland}

\date{13.01.2021}

\begin{abstract}
We analyze the two-orbital (with $d_{x^2-y^2}$ and $d_{3z^2-r^2}$ orbitals) analogue of the $t$-$J$-$U$ model as applied to the description of the copper-oxide monolayer deposited on the Bi$_2$Sr$_2$CaCu$_2$O$_{8+\delta}$ substrate (CuO$_2$/BSCCO). We show, that an $extended$ $s$-$wave$ superconducting dome appears in the heavily hole overdoped regime of the model, with dominant contribution to the pairing coming from the $d_{3z^2-r^2}$ orbitals. Also, the lower critical doping for the appearance of the SC state concurs with the Lifshitz Transition after which the hole-like Fermi pockets are created around the $M$ points in the Brillouin zone. The obtained results are in accord with the recent experimental result [cf. Y. Zhong et al., Sci. Bull. {\bf 61}, 1239 (2016)]. An analogous two-band description is also analyzed in the context of the  Ba$_2$CuO$_{4-y}$ bulk compound, where the heavily hole-overdoped situation is believed to be reached [cf. W. M. Li et al. PNAS {\bf 116}, 12156 (2019)]. As we show, a two dome structure can be realized in such system with the $d$- and $s$-$wave$ symmetries of the gap corresponding to the low hole-doping and heavily hole overdoped cases, respectively.
\end{abstract}

\maketitle
\section{Introduction}

In recent years the fabrication of a monoatomic CuO$_2$ layer deposited on Bi$_2$Sr$_2$CaCu$_2$O$_{8+\delta}$ substrate (CuO$_2$/BSCCO) has been achieved, with a robust U-shaped local density of states (LDOS), which is characteristic for a nodeless superconducting (SC) gap\cite{Zhong2016}. Such finding is in contradiction with the situation known from the high-T$_C$ superconducting cuprates, where the copper-oxide planes, stacked inside a bulk material, are believed to be responsible for the appearance of the planar $d$-$wave$ (nodal) paired state\cite{Harligen1995,Ogata_2008}. So far, it is not completely resolved why those two distinct superconducting states can be observed in so similar systems and, if the Cooper pair creation in CuO$_2$/BSCCO has intrinsic or extrinsic origin. Under such circumstances, the research oriented on the CuO$_2$/BSCCO system has attracted much attention, since it can cast new light on the long lasting fundamental problem concerning the character of SC pairing in the high-$T_C$ copper-based compounds.

It has been argued that the pairing in CuO$_2$/BSCCO results from the proximity effect, with the $d$-$wave$ nodal lines avoided due to specific Fermi-surface topology in the monolayer or caused by magnetic ordering\cite{Zhu2016,Wang2017,Wang2018,Zhu2017}. On the other hand, an intrinsic spin-orbital exchange induced pairing mechanism has been proposed, which leads to the nodeless superconducting state in a straightforward manner\cite{Jiang2018}. What distinguishes the CuO$_2$/BSCCO system from the well known cuprates is that a highly hole-overdoped regime can be reached due to the charge transfer from the monolayer to the BSCCO substrate, which takes place in order to maintain charge neutrality\cite{Jiang2018}. It should be noted, that such high hole-doping levels can also be obtained as a result of the the high-pressure oxidized synthesis of the bulk system Ba$_2$CuO$_{4-y}$ (Ba214) in which superconductivity has been reported very recently\cite{Li12156,Li85}, as well as in Sr$_2$CuO$_{4-\delta}$, discussed already some time ago\cite{Liu2006}. 

According to the DFT calculations, in the CuO$_2$/BSCCO system, the 3d$^8$ configuration is reached with both $d_{x^2-y^2}$ and $d_{3z^2-r^2}$ orbitals becoming active suggesting that a minimal two-band model description may be appropriate\cite{Jiang2018}. A similar approach has been applied to the Ba214 compound, where additionally a doping dependent energy splitting between the two  active orbitals has been taken into account, which changes between Cu$^{2+}$ and Cu$^{3+}$ configurations as the system evolves from the zero-doping to the heavily hole-overdoped situation, respectively\cite{Maier2019}. It should be noted that such two-band structure, appearing in both CuO$_2$/BSCCO and Ba214, is significantly different from that discussed earlier within the single band\cite{Zegrodnik2017,Zegrodnik2017_2} or the three-band description\cite{Zegrodnik2019} of the Cu-O planes in the previously known bulk cuprates. In the latter, the hybridized band with the main contribution coming from the $d_{x^2-y^2}$ states is well separated in energy from the two lower lying bands and a single Fermi surface appears (cf. Fig. 2 in Ref. \onlinecite{Zegrodnik2019}). Here, in the the heavily hole overdoped regime of CuO$_2$/BSCCO or Ba214, we have a degenerate situation in which the Fermi level crosses both bands (cf. Sec. II below). In such case, the additional interactions have to be taken into account represented by the interorbital Coulomb-repulsion and the Hund's rule exchange terms. Also, the resulting Fermi surface topology can affect the energetically favorable pairing symmetry. This circumstance leads to a significant extension of the earlier study of the correlated electrons in the cuprates and constitutes the motivation for analyzing the potentially different superconducting features of the seemingly very similar physical system.

In this paper, we propose a two-band analogue of the $t$-$J$-$U$ model as applied to the description of both CuO$_2$/BSCCO and Ba214. As we have shown in recent years, the single-band $t$-$J$-$U$ approach leads to a very good quantitative agreement between theory and experiment for selected principal features of the bulk cuprates\cite{Zegrodnik2017,Zegrodnik2017_2}. Here, by going along similar lines, we allow for nonzero multiple occupancies and, at the same time, take into account the intersite kinetic exchange interaction term. The former are tuned by the presence of the significant intra-atomic Coulomb repulsion, while the latter contributes to the realization of the real-space pairing scenario in the system. It is important to see if the principal features of the superconducting state in the CuO$_2$/BSCCO and Ba214 can be modeled within such theoretical approach, quite similar to the one applied earlier to the well known cuprates. To take into account the correlation effects we apply the diagrammatic expansion of the variational wave function (DE-GWF) method and analyze in systematically the evolution of the paired state across the phase diagram. We discuss in detail our results in the view of the recent theoretical hypothesis that an $extended$ $s$-$wave$ superconducting dome might appear as a function of hole doping in the CuO$_2$/BSCCO system\cite{Jiang2018}. For the case of the newly discovered Ba214 copper-based compound we analyze a wide doping range, reaching the zero doping on the one side and the heavily hole overdoped situation on the other. Our results suggest that the appearance of the $extended$ $s$-$wave$ pairing is indeed possible in the latter regime.



\section{Model and method}
\begin{figure}
 \centering
 \includegraphics[width=0.5\textwidth]{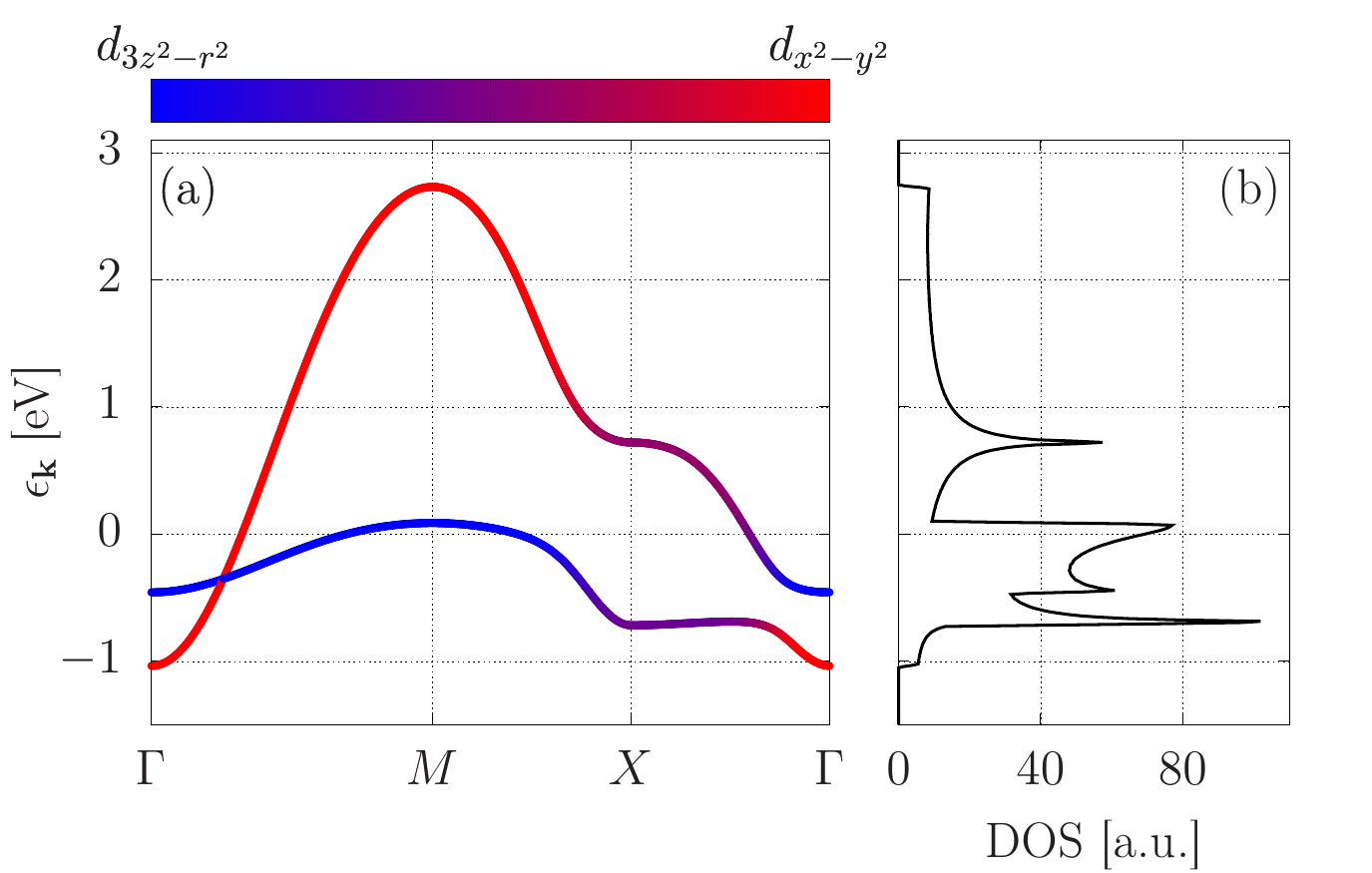}
 \caption{The band structure resulting from the Hamiltonian $\hat{H}_{TBA}$ (a), together with the corresponding density of states (b). The level $\epsilon_{\mathbf{k}}=0$ corresponds to the Fermi energy, which crosses both bands for $\delta=1$ hole doping . The orbital content of the dispersions is marked by the colored scale. Note that this figure correspond to $H_{TBA}$ only, without all the Coulomb and exchange interaction terms which will modify the presented picture.}
 \label{fig:bands}
\end{figure}

We model the CuO$_2$ monolayer on the BSCCO substrate with the use the Hamiltonian consisting of the kinetic energy part ($\hat{H}_{TBA}$), the inter-site exchange part ($\hat{H}_{J}$), and the onsite interaction part ($\hat{H}_{I}$),
\begin{equation}
    \hat{H}=\hat{H}_{TBA}+\hat{H}_{J}+\hat{H}_I,
    \label{eq:Hamiltonian_general}
\end{equation}
where the first term represents the bare-band structure within the tight binding approximation (TBA) with the two active  $d_{x^2-x^2}$ and $d_{3z^2-r^2}$ orbitals and a hybridization between them

\begin{equation}
 \hat{H}_{TBA}=\sum_{\mathbf{k}ll'\sigma}\epsilon^{ll'}_{\mathbf{k}}\hat{c}^{\dagger}_{\mathbf{k}l\sigma}\hat{c}_{\mathbf{k}l'\sigma}+\sum_{\mathbf{k}l\sigma}(\epsilon^l_0-\mu)\hat{n}_{\mathbf{k}l\sigma}
 \label{eq:Hamiltonian_start}
\end{equation}
where $\hat{c}^{\dagger}_{\mathbf{k}l\sigma}$ ($\hat{c}_{\mathbf{k}l\sigma}$) are the creation (anihilation) operators of electrons with momentum $\mathbf{k}$, spin $\sigma$ and orbital index $l=x,\;z$ corresponding to $d_{x^2-y^2}$ and $d_{3z^2-r^2}$ orbitals, respectively, placed on a square lattice. The explicit form of the dispersion relations are provided below
\begin{equation}
    \begin{split}
        \epsilon_{\mathbf{k}}^{ll}=&-2t_l(\cos k_x+\cos k_y)-4t_l'\cos k_x\cos k_y\\
        &-2t_l''(\cos 2k_x+\cos 2k_y)\\
        \epsilon_{\mathbf{k}}^{xz}=&2t_{xz}(\cos k_x-\cos k_y)\\
        &+2t_{xz}''(\cos 2k_x-\cos 2k_y),\\
    \end{split}
\end{equation}
where the values of the hopping parameters have been taken from Ref. \onlinecite{Jiang2018} and are listed in Tab. \ref{tab:hopping_par_1} for the sake of clarity. The second term in $\hat{H}_{TBA}$ introduces the onsite (atomic) energies and the chemical potential $\mu$. The atomic energies are set to: $\epsilon^{x}_0=0.0\;$eV and $\epsilon^{z}_0=-0.91\;$eV and represent the energy splitting between the two orbitals of the model. The resulting band structure is provided in Fig. \ref{fig:bands}, together with the corresponding density of states (DOS). In general, the doping is defined as $\delta\equiv 3-n$, where $n$ is the total number of electrons on a lattice site. Therefore, the extremely hole overdoped situation corresponds to $\delta\lesssim 1$ in which the Fermi energy crosses both bands as shown in Fig. \ref{fig:bands} (a). It should be emphasized that the situation presented in Fig. \ref{fig:bands} corresponds to the $\hat{H}_{TBA}$ Hamilotnian only, without all the exchange and interaction terms contained in $\hat{H}_{J}$ and $\hat{H}_{I}$. The effects resulting from the latter two factors will modify the electronic structure, making it hole doping dependent. Nevertheless, the appearance of a Lifshitz transition (LT), as the system evolves towards the extremely hole-overdoped regime ($\delta\lesssim 1$), is present also after the inclusion of $\hat{H}_{J}$ and $\hat{H}_{I}$.

\begin{table}
 \caption{The TBA model parameters for CuO$_2$/BSCCO corresponding to the 1st, 2nd, and 3rd nearest neighbor intra-orbital ($x$ and $z$) and inter-orbital ($xz$) hoppings in eV. Values of the parameters have been taken from Ref. \onlinecite{Jiang2018}. }
\begin{center}
\begin{tabular}{ c|c|c|c } 
 \hline\hline
   & $t_x$ & $t_z$ & $t_{xz}$ \\ 
 \hline
 1st ($t$) & $0.47$ & $0.0682$ & $0.178$ \\ 
 \hline
  2nd ($t'$) & $-0.0932$ & $0.0109$ & $0.0$ \\ 
\hline
  3rd ($t''$) & $0.0734$ & $0.0$ & $0.0258$ \\ 
 \hline\hline
 \end{tabular}
 \label{tab:hopping_par_1}
\end{center}
\end{table}

The intersite exchange interaction terms appearing in our Hamiltonian are as follows
\begin{equation}
\begin{split}
 \hat{H}_{J}&=J\sum_{ijl}\mathbf{\hat{S}}_{il}\mathbf{\hat{S}}_{jl}-J'\sideset{}{''}\sum_{ijll'}\hat{c}^{\dagger}_{il\uparrow}\hat{c}^{\dagger}_{jl\downarrow}\hat{c}_{il'\downarrow}\hat{c}_{jl'\uparrow},
 \end{split}
 \label{eq:Hamiltonian_pairing}
\end{equation}
where $J$ is the kinetic exchange integral and $\{\mathbf{\hat{S}}_{il}\}$ are the spin-1/2 operators. Such term leads to a spin-singlet real-space pairing and appears both in the so called $t$-$J$ and $t$-$J$-$U$ models extensively discussed in the context of the SC state of the bulk cuprates\cite{Ogata_2008,Zegrodnik2017,Zegrodnik2017_2}. Here, we take $J=0.2\;$eV, which has been chosen so as to obtain a realistic value of the measured nodeless spectral gap in CuO$_2$/BSCCO which is $\approx 20\;$meV\cite{Zhong2016}. An analogue of the $t$-$J$ model for the case of the two-band description of the CuO$_2$/BSCCO system has been derived in Ref. \onlinecite{Jiang2018}, as well as discussed in Ref. \onlinecite{le2019electronic} in the context of the Ba214 cuprate. In such approach, in addition to the $\sim J$ term, also an inter-band pair-hopping and inter-band pairing terms appear. In general, a significant wave vector mismatch between the two Fermi surface sheets would lead to a strong suppression of the Cooper pairing between electrons from different bands with opposite spins and momenta, as discussed for a slightly different model in Ref. \onlinecite{Zegrodnik_fflo}. Therefore, the inter-band pairing terms can be safely omitted here, leaving only the kinetic exchange $\sim J$ and the pair-hopping term $\sim J'$. In our calculations we have taken $J'$ one order of magnitude smaller than $J$. In contrast to the situation analyzed in Ref. \onlinecite{Jiang2018}, where the multiple occupancies are projected out, here, we allow for nonzero multiple occupancies, which are going to be tuned by the presence of the Coulomb repulsion terms. As it has been shown, the presence of small but non-zero multiple occupancies, together with the the exchange interaction term allows for the reconstruction of selected principal features of the superconducting state in the bulk cuprates within the single-band $t$-$J$-$U$ model\cite{Zegrodnik2017,Zegrodnik2017_2}. Therefore, along the same lines, we include the intra-site interaction part, $H_I$, alongside with $H_J$. The explicit form of the $H_I$ term is as follows
\begin{equation}
\begin{split}
 \hat{H}_{I}=&U\sum_{il}\hat{n}_{il\uparrow}\hat{n}_{il\downarrow}+\big(V-\frac{1}{2}J_H\big)\sideset{}{''}\sum_{ill'}\hat{n}_{il}\hat{n}_{il'}\\
 &-J_H\sideset{}{''}\sum_{ill'}\mathbf{\hat{S}}_{il}\cdot\mathbf{\hat{S}}_{il'}
 +J_H\sideset{}{''}\sum_{ill'}\hat{c}^{\dagger}_{il\uparrow}\hat{c}^{\dagger}_{il\downarrow}\hat{c}_{il'\downarrow}\hat{c}_{il'\uparrow}\\
 \end{split}
 \label{eq:Hamiltonian_U}
\end{equation}
where $U$ and $V$ are the intra- and inter-orbital Coulomb repulsion integrals, respectively, whereas the terms $\sim J_H$ account for the interorbital Hund's-rule contribution. The double primed summations correspond to taking $l\neq l'$ only. It should be noted, that since $J_H$ is one order of magnitude smaller than $U$ or $V$, and as we ignore the intra-site paired state, as well as the spin-flip hopping processes, the last term in (\ref{eq:Hamiltonian_U}) introduces negligibly small contribution to the system energy and has been disregarded here. In our calculations we take the relation typical for the $e_g$ complex: $U=V+2J_H$ and $J_H=0.1\;U$ with $U=1.5\;$eV.

To take into account the appearance of the Coulomb repulsion of a significant magnitude, we apply the diagrammatic expansion of the Gutzwiller wave function (DE-GWF) method. The generalized Gutzwiller-type projected many-particle wave function is taken in the form
\begin{equation}
 |\Psi_G\rangle\equiv\hat{P}|\Psi_0\rangle\equiv\prod_{il}\hat{P}_{il}|\Psi_0\rangle \;,
 \label{eq:GWF}
\end{equation}
where $|\Psi_0\rangle$ represents the wave function of uncorrelated state to be selected self-consistently later. The projection operator has the following form
\begin{equation}
 \hat{P}_{il}\equiv \sum_{\Gamma}\lambda_{\Gamma|il}|\Gamma\rangle_{il\;il}\langle\Gamma|\;,
 \label{eq:P_Gamma}
\end{equation}
where $\lambda_{\Gamma|il}$ are the variational parameters determining relative weights of the corresponding $|\Gamma\rangle_{il}$ states, which in turn represent the local basis configurations on the atomic sites with the two types of orbitals ($l\in\{{d_{x^2-y^2},d_{3z^2-r^2}}\}$). Explicitly,
\begin{equation}
|\Gamma\rangle_{il}\in \{|\varnothing\rangle_{il}, |\uparrow\rangle_{il}, |\downarrow\rangle_{il},
|\uparrow\downarrow\rangle_{il}\}\;,
\label{eq:local_states}
\end{equation}
where the consecutive states represent the empty, singly, and doubly occupied local configurations, respectively. As can be seen , the variational parameters, which adjust the weight of the local electronic configurations appearing in the many-particle wave function, are explicitly orbital-dependent. The DE-GWF scheme allows to express, in a relatively compact manner, the energy of the system in the $|\Psi_G\rangle$ state via
\begin{equation}
    \langle\hat{H}\rangle_G=\frac{\langle\Psi_G|\hat{H}|\Psi_G\rangle}{\langle\Psi_G|\Psi_G\rangle}\equiv\frac{\langle\Psi_0\hat{P}|\hat{H}|\hat{P}\Psi_0\rangle}{\langle\Psi_0|\hat{P}^2|\Psi_0\rangle},
\end{equation}
in terms of the uncorrelated hopping and pairing expectation values $P_{ijll'}=\langle\hat{c}^{\dagger}_{il\sigma}\hat{c}_{jl'\sigma}\rangle_0$, $S_{ijl}=\langle\hat{c}^{\dagger}_{il\uparrow}\hat{c}^{\dagger}_{jl\downarrow}\rangle_0$, as well as the variational parameters $\{\lambda_{\Gamma}\}$. In the next step the grand-canonial potential $\mathcal{F}=\langle \hat{H}\rangle_G-\mu_Gn_G$ is minimized, where $\mu_G$ and $n_G$ are respectively the chemical potential and the number of particles per lattice site determined in the correlated state. Such a procedure reduces the number of configurations which lead to the increased interaction energies in the correlated state. Details of the DE-GWF scheme as applied to the analyzed here two-band approach are deferred to Appendix A. Moreover, selected applications of the method are provided in Refs. \onlinecite{Zegrodnik2017, Zegrodnik2019, Wysokinski2016}. 

The parameters, which are typically analyzed when studying the SC state within the variational wave function-based approaches, are the so-called correlated pairing amplitudes defined in real-space as $\Delta^G_{ij|l}\equiv\langle\hat{c}^{\dagger}_{il\uparrow}\hat{c}^{\dagger}_{jl\downarrow}\rangle_G$. We also exhibit here the so-called effective gap parameters which are in the units of energy and can be associated with the spectral gaps opening in the band structure, $\Delta^{\mathrm{eff}}_{ij|l}=\partial\mathcal{F}/\partial S_{ijl}$\cite{Zegrodnik2017,Zegrodnik2017_2}. In this work, we consider both the $extended$ $s$-$wave$ and $d$-$wave$ pairings, which impose the following conditions on $\Delta^G_{ijl}$ and $\Delta^{\mathrm{eff}}_{ijl}$
\begin{equation}
    \begin{split}
        \Delta^G_{(\pm 1,0)|l}&=\Delta^G_{(0,\pm 1)|l}\equiv \Delta^G_{s|l},\\
        \Delta^G_{(\pm 1,0)|l}&=-\Delta^G_{(0,\pm 1)|l}\equiv \Delta^G_{d|l},\\
        \Delta^{\mathrm{eff}}_{(\pm 1,0)|l}&=\Delta^{\mathrm{eff}}_{(0,\pm 1)|l}\equiv \Delta^{\mathrm{eff}}_{s|l},\\
        \Delta^{\mathrm{eff}}_{(\pm 1,0)|l}&=-\Delta^{\mathrm{eff}}_{(0,\pm 1)|l}\equiv \Delta^G_{d|l},\\
    \end{split}
    \label{eq:d_s_wave_conditions}
\end{equation}
where $\Delta_{\mathbf{R}_{ij}|l}$ depend only on relative distance $\mathbf{R}_{ij}\equiv\mathbf{R}_i-\mathbf{R}_j$ due to the assumed homogeneity of the system and $\mathbf{R}_{ij}$ is expressed in units of lattice parameter $a$. Here we analyze only the nearest-neighbor pairings as they are the dominant contributions to the paired state, in the same manner as in other models of the same class\cite{Kaczmarczyk_2014,Zegrodnik2017_2,Zegrodnik2019}. Nevertheless, both pairings and hoppings up to the fourth nearest-neighbor are taken into account. An important parameter of the DE-GWF calculations is the so-called order of the diagrammatic expansion (defined in Appendix A). It has been shown that the 3-5 order calculations lead to sufficient accuracy of the method\cite{Bunemann2012, Kaczmarczyk_2014}. The results presented in Sec. III have been obtained within the third order expansion calculations.

\section{Results}

\subsection{Nodeless superconducting dome in CuO$_2$/BSCCO}

\begin{figure}
 \centering
 \includegraphics[width=0.5\textwidth]{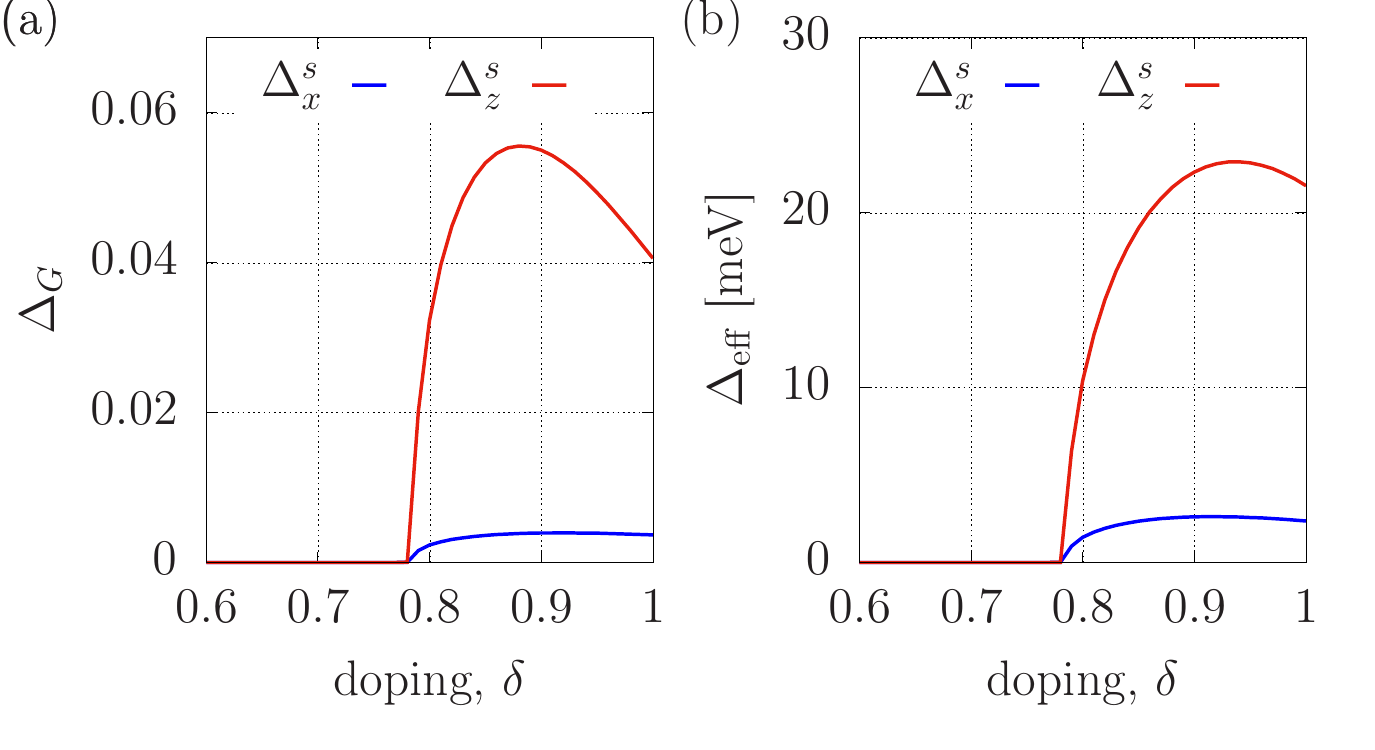}
 \caption{The correlated (a) and the effective (b) gap parameters corresponding both to $x$ and $z$ orbitals plotted as a function of hole doping for the TBA parameters corresponding to the CuO$_2$/BSCCO system.}
 \label{fig:delt_1}
\end{figure}

\begin{figure}
 \centering
 \includegraphics[width=0.5\textwidth]{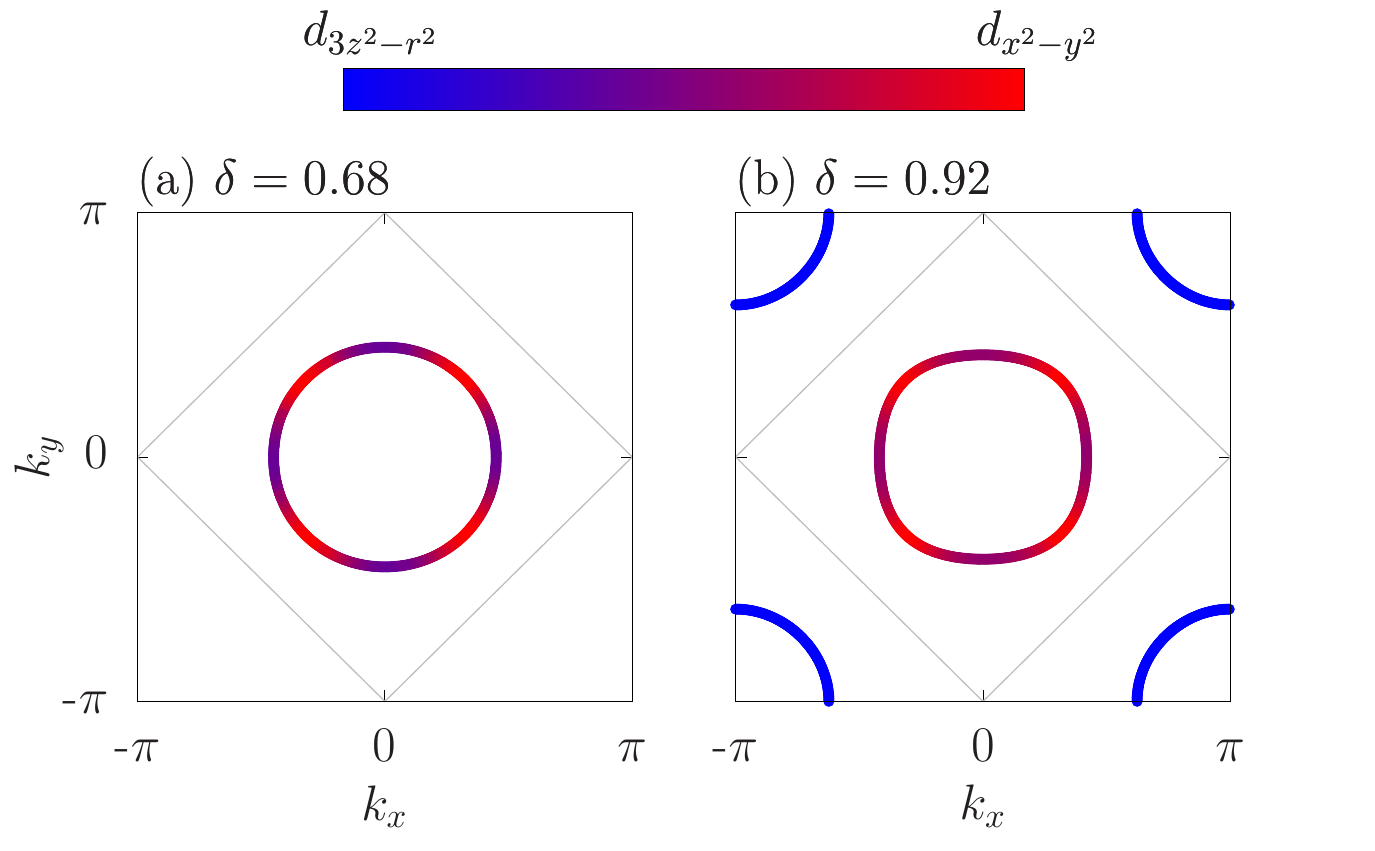}
 \caption{The Fermi surface topology for two selected values of the hole doping: $\delta=0.68$ (a), $\delta=0.92$ (b). In (a) $extended$ $s$-$wave$ paired states appear, hence, the so-called underlying Fermi surfaces have been drawn. In (b) SC state is suppressed and only normal state becomes stable (cf. Fig. \ref{fig:delt_1}). The $extended$ $s$-$wave$ symmetry nodal lines are marked by grey solid line.}
 \label{fig:fermi_1}
\end{figure}

In Fig. \ref{fig:delt_1} we show the hole doping dependence of both the correlated and the effective gap amplitudes, $\Delta_{G}$ and $\Delta_{\mathrm{eff}}$, corresponding to $x$ and $z$ orbitals of the CuO$_2$/BSCCO monolayer. Note that the gap $\Delta_{\mathrm{eff}}$ is in physical units, whereas the correlated gap $\Delta_G$ is dimensionless and represents the off-diagonal correlation function only. As one can see, in the heavily-hole doped situation ($\delta\lesssim 1$) a region of stability of the $extended$ $s$-$wave$ superconducting state appears with the maximal values of the effective gap $\approx 20$ meV, which agrees with the experimental result reported in Ref. \onlinecite{Zhong2016}. The $extended$ $s$-$wave$ pairing between the particles occupying the $z$ orbitals dominates over the corresponding $x$-orbital pairing and the lower critical doping value for the appearance of the $extended$ $s$-$wave$ SC state corresponds to the Lifshitz transition (at $\delta_c\approx 0.8$). For $\delta>\delta_c$ the lower band, which consists mainly of the $z$ states (cf. Fig. \ref{fig:bands}), becomes partially filled, leading to the creation of an additional Fermi surface sheet. In Fig. \ref{fig:fermi_1} we show explicitly the evolution of the Fermi surface topology as one passes the LT, with those additional Fermi surface sheets consisting mainly of the $d_z$ states seen in (b). One should note that the $extended$ $s$-$wave$ nodal lines lie in between the two underlying Fermi surface sheets shown in Fig. \ref{fig:fermi_1} (b), leading to a realization of the so-called $s\pm$ paired state.



\begin{figure}
 \centering
 \includegraphics[width=0.5\textwidth]{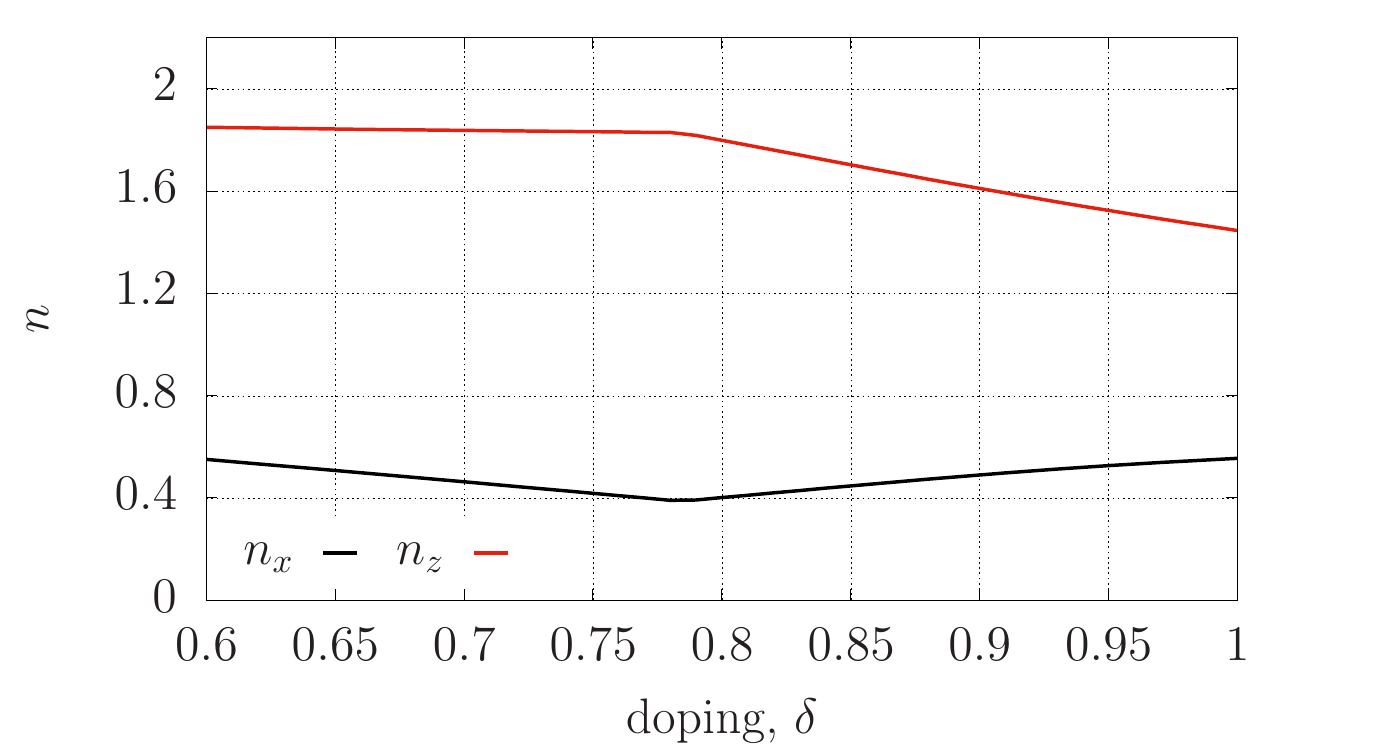}
 \caption{Number of particles per atomic site occupying the $x$ and $z$ orbitals as a function of hole doping for the TBA parameters corresponding to the CuO$_2$/BSCCO system (listed in Table \ref{tab:hopping_par_1}).}
 \label{fig:n_tot}
\end{figure}

In Fig. \ref{fig:n_tot}, we show the number of electrons per lattice site partitioned between the two orbitals ($n=n_x+n_z$), as a function of hole doping. As one can see, the Lifshitz transition is seen also here where the change of monotonicity of $n_x$ appears after the lower band becomes partially filled with increasing hole doping. This effect is induced by the presence of the inter-orbital Coulomb repulsion and can be understood even within a simple mean-field (MF) picture. Namely, according to the MF treatment, due to the $\sim V$ term a doping dependent contributions to the orbital-energy levels appears, which are: $V\langle n_{z}\rangle$ and $V\langle n_{x}\rangle$ for $x$ and $z$ orbitals, respectively. After the LT, $n_z$ decreases significantly, since the lower band becomes partially occupied. This leads to a decrease of the $x$-orbital energy level due to the V$\langle n_{z}\rangle$ term. The latter, in turn, results in electron transfer to the $x$ orbital. That is why the decrease of $n_z$ after the LT leads to an increase of $n_x$. Similar effect has also been reported both theoretically and experimentally for LAO/STO\cite{Zegrodnik_2020,maniv2015strong,Smink}. 

\subsection{Two superconducting domes in Ba214}
\begin{figure}
 \centering
 \includegraphics[width=0.5\textwidth]{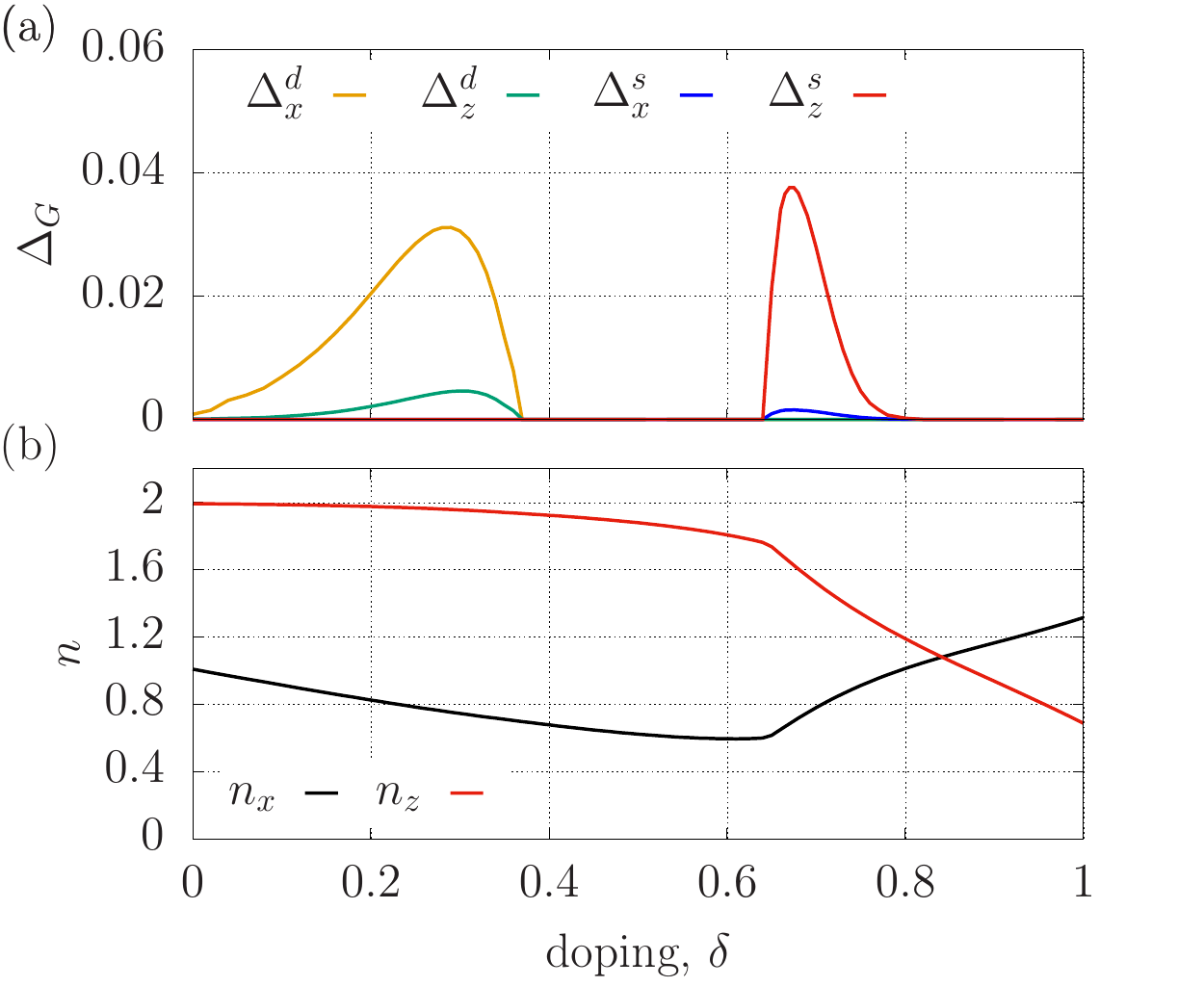}
 \caption{(a) Correlated amplitudes $\Delta_G$ corresponding both to $x$ and $z$ orbitals as a function of hole doping doping, $\delta$ and for the model parameters corresponding to the Ba214 compound. (b) The number of particles per atomic site occupying the $x$ and $z$ orbitals as a function of hole doping. Figure (a) represents the phase diagram of the ground-state superconducting states.}
 \label{fig:delt_2}
\end{figure}

\begin{figure}
 \centering
 \includegraphics[width=0.5\textwidth]{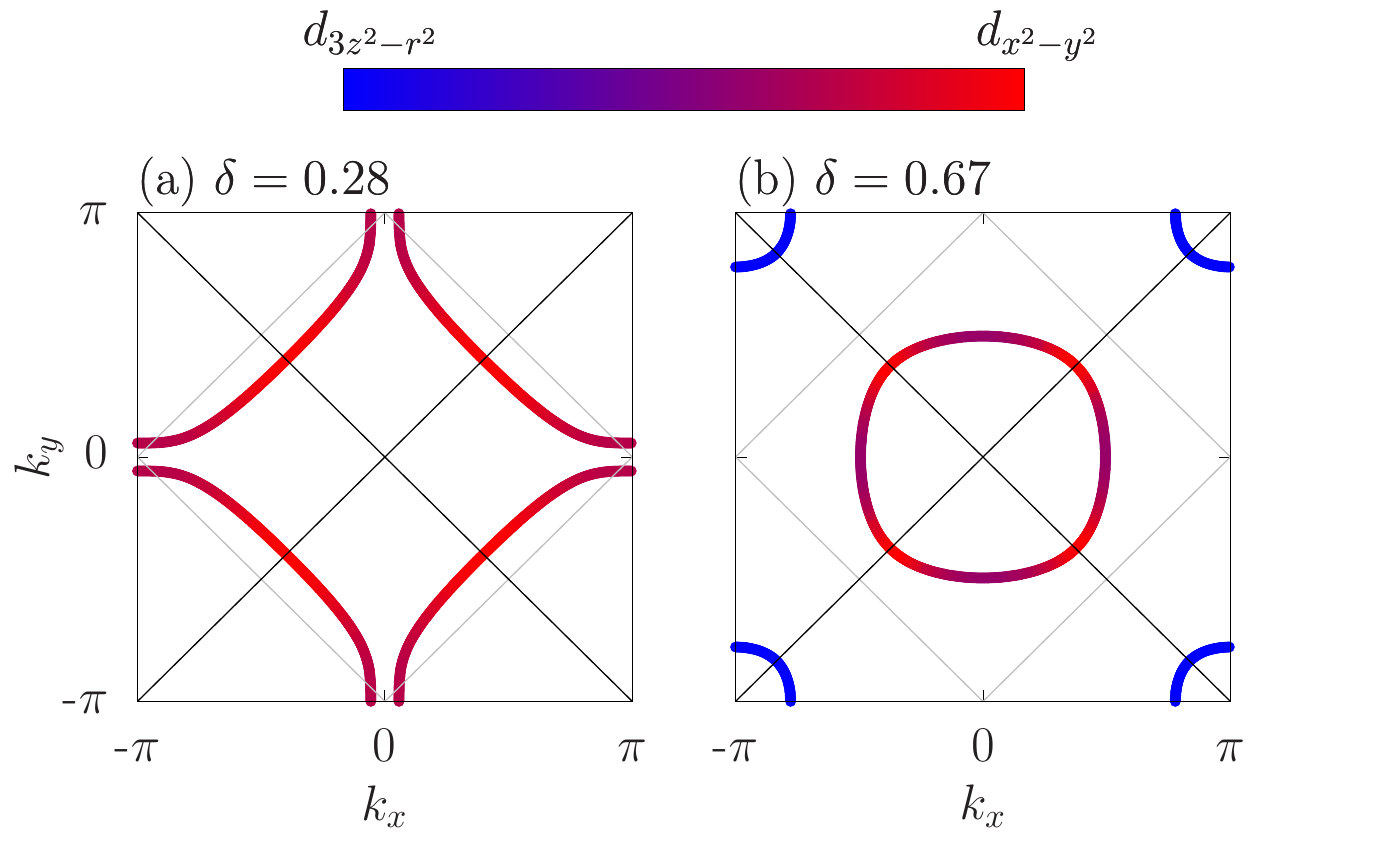}
 \caption{The underlying Fermi surfaces for two selected values of the hole doping: $\delta=0.28$ (a), $\delta=0.67$ (b), which correspond to the appearance of $d$- and $s$- wave superconducting domes, respectively (cf. Fig. \ref{fig:delt_2}).  The $d$- and $s$-$wave$ nodal lines are marked by black and grey solid lines, respectively.}
 \label{fig:fermi_h}
\end{figure}

Since it is believed that a heavily hole doped situation can also be realized in the newly discovered bulk superconductor Ba214, we have applied our approach for the two-band description of that system. We have analyzed the evolution of the paired state throughout the doping region, reaching the zero doped case on the one side ($\delta=0$) and the extremely heavily hole doped regime on the other ($\delta=1$). Following Maier et al.\cite{Maier2019} we have taken into account the doping dependent energy splitting between $x$ and $z$ orbitals. The tight binding hopping parameters corresponding to Ba214 are provided in Table \ref{tab:hopping_par_2} while the atomic energies are: $\epsilon^x_0=-0.222\;$eV and $\epsilon^z_0=0.661-5(1-\delta)\;$eV, with the latter being doping dependent\cite{Maier2019}. As we show in Fig. \ref{fig:delt_2}, for such band structure, our approach leads to the two-dome behavior of the pairing amplitudes. For $\delta\gtrsim 0$, where the $d$-$wave$ paired state is stable, we have a hole-like underlying Fermi surface, composed mainly of the $d_{x}$ states [cf. Fig. \ref{fig:fermi_h} (a)]. This situation is in close connection to the one well known bulk cuprates. As one can see, the FS shown in Fig. \ref{fig:fermi_h} (a) is in close proximity to the $extended$ $s$-$wave$ nodal lines at its full extent, making it energetically unfavorable for this pairing symmetry to appear. For large values of $\delta$ the Fermi surface topology changes significantly [Fig. \ref{fig:fermi_h} (b)] and the $extended$ $s$-$wave$ paired phase becomes stable similarly as in CuO/BSCCO. However, this time the stability regime of the latter is visibly narrower and the upper critical doping coincides with the band inversion seen in Fig. \ref{fig:delt_2} (b). The latter is due to the $\delta$-dependent energy splitting between $x$ and $z$ orbitals, which also enhances the monotonicity change of $n_x$ seen after the Lifshits transition has taken place. One should note, that for the case of the $d$-$wave$ symmetry, the pairing between the $x$ orbitals is the dominant, in contrast to the $extended$ $s$-$wave$ pairing.

Interestingly, our result differs from the one presented in Ref. \onlinecite{Maier2019}, where the RPA treatment is applied. Namely, in the heavily hole doped regime we have obtained only $extended$ $s$-$wave$ SC solution, whereas both $s$- and $d$-$wave$ pairings of comparable magnitudes are visible in the spin-fluctuation strength within the RPA calculations. In connection to that, a close competition between the two pairing symmetries, with the $s$-$wave$ dominating in the $\delta=1$ regime, has been studied within the Lieb lattice model of Ba214 in in Ref. \onlinecite{Yamazaki2020}.

\begin{table}
 \caption{The TBA model parameters of the Ba214, corresponding to the 1st, 2nd, and 3rd nearest neighbor intra-orbital ($x$ and $z$) and inter-orbital ($xz$) hoppings in eV. Values of the parameters have been taken from Ref. \onlinecite{Maier2019}.}
\begin{center}
\begin{tabular}{ c|c|c|c } 
 \hline\hline
   & $t_x$ & $t_z$ & $t_{xz}$ \\ 
 \hline
 1st ($t$) & $0.504$ & $0.196$ & $0.302$ \\ 
 \hline
  2nd ($t'$) & $-0.067$ & $0.026$ & $0.0$ \\ 
\hline
  3rd ($t''$) & $0.13$ & $0.029$ & $0.051$ \\ 
 \hline\hline
 \end{tabular}
 \label{tab:hopping_par_2}
\end{center}
\end{table}

\section{Conclusions and outlook}

We have proposed a two-band analogue of the $t$-$J$-$U$ model for the description of both the CuO$_2$/BSCCO system, as well as bulk Ba214 compound. We have shown that both nodal ($d$-$wave$) and nodeless ($extended$ $s$-$wave$) superconducting states can be obtained within a single picture, in which the intersite kinetic exchange interaction is responsible for the appearance of the real-space spin-singlet pairing. 

According to our study, the appearance of the $extended$ $s$-$wave$ pairing in the heavily hole overdoped regime is associated with the hole-like Fermi surface sheets centered at the $M$ points of the Brillouin zone, which are created after the Lifshitz transition has taken place when increasing the hole doping. This effect could be verified experimentally for the bulk Ba214 compound by determining if the lower critical doping for the $extended$ $s$-$wave$ paired state correlates with the LT transition point, determined, e.g., from ARPES measurements. For CuO$_2$/BSCCO, the LT can be identified in the magneto-transport as the transition point between the one- to two-carrier transport similarly, as has been done in Ref. \onlinecite{joshua2012universal} for the case of LAO/STO interface. Note that the LT may also be of importance when it comes the possible exotic magnetic excitations of the model as emphasized in Ref. \onlinecite{Zhang_2020}. 

For the case of the Ba214 compound we have obtained the two-dome structure consisting of $d$- and $extended$ $s$-$wave$ pairing amplitudes corresponding to the hole underdoped case and the heavily hole overdoped regime, respectively. According to our analysis, the $z$ orbitals play the dominant role for the $extended$ $s$-$wave$ pairing, similarly as for CuO$_2$/BSCCO, whereas for the $d$-$wave$ pairing, particles on the $x$ orbitals are more important. Due to the doping dependent interband splitting, which appears in the model of Ba214, the band inversion takes place and the $s$-$wave$ pairing regime is narrowed down with respect to the CuO$_2$/BSCCO case.

Another effect analyzed here is the change of monotonicity of the average number of particles occupying the $x$ orbital which takes place at the Lifshitz transition. This is due to the influence of the interorbital Coulomb repulsion term, but is additionally enhanced by the doping dependent energy splitting between the two orbitals, appearing in the model describing the Ba214 compound. Perhaps, such an effect could be verified experimentally by using the NMR technique for the bulk Ba214 or by carrying out electron transport measurements in a Hall bar architecture\cite{joshua2012universal,Biscaras} for the CuO$_2$/BSCCO system . 

It should be noted that here we analyze the results when, kinetic exchange (with $J=0.2\;$eV) dominates over the Hund's-rule coupling ($J_H=0.15\;$eV). In effect, we consider only the situation with the spin-singlet pairing disregarding the spin-triplet contribution\cite{Zegrodnik_triplet}, which becomes predominant in the Hund's metal regime

In summary, we have shown that the variational approach, applied earlier to the cuprates and based on a real-space pairing for correlated particles, can also describe the principal ground-state properties of CuO$_2$/BSCCO, as well as Ba214 bulk compounds. The question is whether such a two-band model can also describe quantum spin and charge excitations, in a simple analogy to the one-band cuprates\cite{Fidrysiak2020}. They may show a mixed ferro- and antiferro- character depending on a concrete system.

\section{Acknowledgement}
This work was supported by National Science Centre, Poland (NCN) according to decision 2017/26/D/ST3/00109 and in part by PL-Grid Infrastructure. JS acknowledges the financial support by the Grant OPUS No. UMO-2018/29/B/ST3/02646 from the National Science Centre (NCN), Poland.

\appendix
\section{}
The main task within the DE-GWF calculation scheme is to derive a relatively compact analytical expression for the system energy expectation value in the variational correlated state given by Eq. (\ref{eq:GWF}). In order to simplify significantly the calculations and improve the convergence one imposes the following condition on the $\hat{P}_{il}$ operator \cite{Bunemann2012,Gebhard1990} 
\begin{equation}
 \hat{P}_{il}^2\equiv 1+\alpha_{il}\hat{d}^{\textrm{HF}}_{il}\;,
 \label{eq:condition}
\end{equation}
where $\hat{d}^{\textrm{HF}}_{il}=\hat{n}_{il\uparrow}^{\textrm{HF}}\hat{n}_{il\downarrow}^{\textrm{HF}}$, $\hat{n}_{il\sigma}^{\textrm{HF}}\equiv\hat{n}_{il\sigma}-n_{l0}$, 
with $n_{l0}\equiv\langle\Psi_0|\hat{n}_{il\sigma}|\Psi_0\rangle$, and $\alpha_{il}$ is yet another variational parameter in addition to $\lambda_{\Gamma|il}$ introduced already in Eq. (\ref{eq:P_Gamma}). By comparing Eqs. (\ref{eq:P_Gamma}) and (\ref{eq:condition})
we can express the parameters $\lambda_{\Gamma|il}$ with the use of $\alpha_{il}$. 
\begin{equation}
\begin{split}
 \lambda^2_{d|l}&=1+\alpha_l(1-n_{l0})^2\\
 \lambda^2_{s|l}&=1-\alpha_ln_{l0}(1-n_{l0})\\
 \lambda^2_{\emptyset|l}&=1+\alpha n_{l0}^2,
 \end{split}
 \label{eq:lambda_x}
\end{equation}
where $\lambda_{\Gamma|l}\in\{\lambda_{\emptyset|l},\lambda_{s|l},\lambda_{d|l }\}$ correspond to states $|\emptyset\rangle_l\;, |\sigma\rangle_l\;, |\uparrow\downarrow\rangle_l$, respectively, and both the site index, $i$, and the spin index, $\sigma$, have been dropped since we are considering a spatially homogeneous, spin-isotropic case. Due to Eq. (\ref{eq:lambda_x}) we are left with one variational parameter per orbital within the procedure.

In the next step, one has to express the expectation values of all the terms appearing in Hamiltonian (\ref{eq:Hamiltonian_start}) in the correlated state $|\Psi_G\rangle$. Here, we show the corresponding expression for the electron hopping term as an example
 \begin{equation}
 \begin{split}
  \langle\Psi_G|&\hat{c}^{\dagger}_{il\sigma}\hat{c}_{jl'\sigma}|\Psi_G\rangle=\\
  &\sum_{k=0}^{\infty}\frac{1}{k!}\sideset{}{'}\sum_{m_1f_1...m_kf_k}\alpha^{k_x}_x\alpha^{k_z}_z
  \langle\tilde{c}^{\dagger}_{il\sigma}\tilde{c}_{jl\sigma}\hat{d}^{\textrm{HF}}_{m_1f_1}...\hat{d}^{\textrm{HF}}_{m_kf_k} \rangle_0\;,
  \end{split}
  \label{eq:diag_sum}
 \end{equation}
where $\hat{d}^{\textrm{HF}}_{\varnothing}\equiv 0$, $\tilde{c}^{(\dagger)}_{il\sigma}\equiv \hat{P}_{il}\hat{c}^{(\dagger)}_{il\sigma}\hat{P}_{il}$ and the index $m$ corresponds to lattice sites, whereas $f$ enumerates the orbitals. The primmed summation on the right hand side is restricted to $(l_h,m_h)\neq (l_{h'},m_{h'})$, $(l_h,m_h)\neq (i,l)$, $(l_h,m_h)\neq (j,l')$ for all $h$, $h'$.
The powers $k_x$ ($k_z$) express how many times the indices $f_h$ on the right hand side of Eq. (\ref{eq:diag_sum})
have the value corresponding to $x$ ($z$) orbital. For given $k$, they fulfill the relation $k_x+k_z=k$.  The maximal $k$, for which the terms in Eq. (\ref{eq:diag_sum}) are taken into account represents the order of calculations. Similar expressions can be derived for the case of the onsite interaction terms contained in (\ref{eq:Hamiltonian_U}) and the intersite exchange interaction terms in (\ref{eq:Hamiltonian_pairing}). However, we do not show them here as they are to lengthy. It has been shown that the first 4-6 terms of the expansion lead to a sufficient accuracy of the method\cite{Bunemann2012, Kaczmarczyk_2014}. The results presented in Sec. III have been obtained within the third order expansion calculations (taking into account first 4 terms).

It should be noted that by carrying out the analogical expansion for all of the terms from Hamiltonian (\ref{eq:Hamiltonian_general}) one expresses the system energy in the correlated state with the use of the variational parameters, $\alpha_{l}$ and the expectation values in the non-correlated state, $\langle...\rangle_0\equiv\langle\Psi_0|...|\Psi_0\rangle$. The latter can be additionally decomposed with the use of Wicks theorem in direct space and written in terms of the so-called hopping and pairing terms $P_{ijll''\sigma}\equiv\langle\hat{c}^{\dagger}_{il\sigma}\hat{c}_{j\sigma} \rangle_0$, $S_{ijll'}\equiv\langle\hat{c}^{\dagger}_{il\uparrow}\hat{c}^{\dagger}_{jl'\downarrow} \rangle_0$, respectively.

To obtain the final form of the wave function, the grand-canonical potential is minimized $\mathcal{F}=\langle\hat{H}\rangle_G-\mu_G\langle\hat{n}\rangle_G$, where $\mu_G$, is the chemical potential. The minimization condition can be cast into the form of the Schr\"odinger equation with the effective Hamiltonian\cite{Kaczmarczyk_2014}
\begin{equation}
\begin{split}
 \hat{\mathcal{H}}_{\textrm{eff}}&=\sideset{}{'}\sum_{ijll'\sigma}t^{\textrm{eff}}_{ijll'}\hat{c}^{\dagger}_{il\sigma}\hat{c}_{jl'\sigma}+\sum_{il\sigma}\epsilon^{\textrm{eff}}_{il}\hat{n}_{il\sigma}\\
 +&\sideset{}{''}\sum_{ijll'}\big(\Delta^{\textrm{eff}}_{ijll'}\hat{c}^{\dagger}_{il\uparrow}\hat{c}^{\dagger}_{jl'\downarrow}+H.c.\big),
 \end{split}
 \label{eq:H_effective}
\end{equation}
where the primed (double primed) summations are restricted to $i\neq j$ ($i\neq j$ and $l\neq l'$) and the effective hopping energies, effective superconducting gap amplitudes, and effective atomic levels are defined in the following manner
\begin{equation}
 t^{\textrm{eff}}_{ijll'}\equiv \frac{\partial\mathcal{F}}{\partial P_{ijll'\sigma}},
 \quad \Delta^{\textrm{eff}}_{ijll'}\equiv \frac{\partial\mathcal{F}}{\partial S_{ijll'}},\quad \epsilon^{\textrm{eff}}_{il}\equiv \frac{\partial\mathcal{F}}{\partial n_{0l}}.
 \label{eq:effective_param}
\end{equation}
In the numerical calculations we take into account the hopping and pairing expectation values up to the fourth nearest neighbor both in the diagrammatic expansion and in the summations appearing in Eq. (\ref{eq:H_effective}). Depending on the symmetry of the gap ($d$- or $s$-$wave$) we impose the proper choice of conditions expressed by Eq. \ref{eq:d_s_wave_conditions}.

The effective Hamiltonian (\ref{eq:H_effective}) can be transformed into the reciprocal space and diagonalized through the 4$\times$4 generalized Bogolubov-de Gennes transformation, which leads to the self-consistent equations for the pairing and hopping expectation values. The minimization over the variational parameters $\alpha_x$ and $\alpha_z$ is incorporated into the  procedure of solving the self-consistent equations. After all the hopping and pairing lines, together with the variational parameters, are determined, one can calculate next the values of superconducting pairing amplitudes between particular sites in the correlated state $|\Psi_G\rangle$. 

\newpage

\bibliography{refs.bib}

\end{document}